# Photo-Switchable Surfactants for Responsive Air-Water Interfaces: Azo vs. Arylazopyrazole Amphiphiles

*Marco Schnurbus[1,2), Richard A. Campbell[3), Jörn Droste[1), Christian Honnigfort[1,2), Dana Glikman[1,2], Philipp Gutfreund[4), Michael Ryan Hansen[1) and Björn Braunschweig[1,2)\**

[1) Institute of Physical Chemistry, Westfälische Wilhelms-Universität Münster, Corrensstraße 28/30, 48149 Münster, Germany

[2) Center for Soft Nanoscience (SoN), Westfälische Wilhelms-Universität Münster, Busso-Peus-Straße 10, 48149 Münster, Germany

[3) Division of Pharmacy & Optometry, University of Manchester, Manchester M13 9PT, United Kingdom

[4) Institut Laue-Langevin, 71 avenue des Martyrs, CS 20156, 38042 Grenoble Cedex 9, France

\*E-mail: braunschweig@uni-muenster.de

**ABSTRACT**

Arylazopyrazoles (AAPs) as substitutes for azo derivatives have gained considerable attention due to their superior properties offering *E*/*Z* photo-isomerization with high yield. In order to compare and quantify their performance, azobenzene tetraethylammonium (Azo-TB) and arylazopyrazole tetraethylammonium (AAP-TB) bromides were synthesized and characterized in the bulk (water) using NMR spectroscopy. At the air-water interface complementary information from vibrational sum-frequency generation (SFG) spectroscopy and neutron reflectometry (NR) has revealed the effects of *E/Z* isomerization in great detail. In bulk water the photostationary states of >89 % for *E/Z* switching in both directions were very similar for the surfactants, while their interfacial behavior was substantially different. In particular, the surface excess $\Gamma$ of the surfactants changed drastically between *E/Z* isomers for AAP-TB (maximum change of $\Gamma$: 2.15 µmol/m²); for Azo-TB the change was only moderate (maximum change of $\Gamma$: 1.02 µmol/m²). Analysis of SFG spectra revealed that strong non-resonant contributions that heterodyned the resonant vibrational bands were proportional to $\Gamma$, enabling the aromatic C-H band to be interpreted as an indicator for changes in interfacial molecular order. Close comparison of $\Gamma$ from NR with the SFG amplitude from the aromatic C-H stretch as a function of concentrations and *E/Z* conformation revealed substantial molecular order changes for AAP-TB. In contrast, only $\Gamma$ and not the molecular order varied for Azo-TB. These differences in interfacial properties are attributed to the molecular structure of the AAP center that enables favorable lateral interactions at the air-water interface, causing closed-packed interfacial layers and substantial changes during *E/Z* photo-isomerization.

# INTRODUCTION

Switchable molecular building blocks for responsive materials have gained considerable attention in the past and are part of a growing field of interest in biochemistry, catalysis, polymer science and in particular colloid and interface chemistry.[1–4] The responsiveness can be reached by different stimuli such as light,[5–14] temperature,[15,16] pH,[17,18] magnetic or electric fields,[19] or through a combination of these factors to yield multiresponsive systems.[3,20–22] In particular, light as a stimulus to trigger interfacial and material responses has the advantage that it can be controlled precisely in time and space, thus offering material control in four dimensions. In order to change the properties of a fluid interface, photoresponsive surfactants that can undergo *E/Z* photo-isomerization reactions have been studied in previous works but were concentrated on azobenzene surfactants for light responsive emulsions,[23–25] organogels,[26] vesicles,[27] microgels[28] and foams.[29–31]

Azobenzenes have a few limitations, however, which restrict their potential use as a photoswitch in many applications: The spectral overlap in optical absorbance of the *E* isomer with that of the *Z* isomer results for most azobenzenes in incomplete photoswitching of photostationary states (PSS) that is <80 % in either direction.[32,33] Furthermore, the *Z* isomer of the azobenzenes shows low thermodynamic stability, and a fast back conversion to the *E* isomer is observed under dark conditions. These factors in fact limit widespread application of azobenzenes and make alternatives like arylazopyrazole[33–35] or spiropyrane[2,36–38] derivatives interesting for current research. One possible research direction is to shift the π → π* band to longer wavelengths or to reach a larger splitting of the n → π* band of the *E* and *Z* isomers from azo derivatives. This splitting was already reported by the groups of Woolley[39,40] and Hecht,[41,42] who have used the ortho-substituted azobenzenes. Woolley at al.[39,40] synthesized azobenzenes with electron-rich substituents in ortho position and showed that it is possible to switch these molecules with red light. Hecht and co-workers[41,42] used ortho-fluoroazobenzenes, which shift the n → π* band to longer wavelengths and show improved photoswitching and increased thermodynamic stability of the *Z* isomer, while Fuchter at al.[34,35] established new light switchable molecules based on arylazopyrazoles derivatives (AAPs), showing a blue shift of the π → π* band and a red shift of the n → π* band. Additionally, AAP derivatives which can be triggered with UV and green light, have superior thermal half-life (~1000 days) and a PSS of >90 %.[33] Clearly, these points demonstrate that the use of AAP moieties can be advantageous compared to azobenzenes. However, while previous studies have made comparisons between AAPs and azobenzenes in the bulk solution[33,34], so far their interfacial properties have not been directly compared.

Here, we introduce two new photoswitchable surfactants that have been synthesized explicitly for this work: a cationic azobenzene tetraethylammonium bromide (Azo-TB) and arylazopyrazole tetraethylammonium bromide (AAP-TB) (Figure 1). These are compared in the bulk via UV/Vis and NMR spectroscopy to determine the PSS and, thus, their bulk performance. At the air-water interface, which we take as a representative platform for a fluid interface, and which has direct functionality both in nature such as with lung surfactants[43,44] as well as in commercial applications such as with foams.[45,46] The light-induced changes are studied using surface tensiometry, vibrational sum-frequency generation (SFG) spectroscopy, and neutron reflectometry (NR). This powerful combination of methods has allowed us previously to reveal a light-induced monolayer to bilayer transition of a AAP sulfonate amphiphiles at the air-water interface[4], which is related to the different molecular structures and amphiphilicity of the corresponding *E* and *Z* isomers. These new insights of the physicochemical changes upon photoswitching at the air-water interface were only possible by making use of the unique structure sensitivity of SFG spectroscopy for non-centrosymmetric environments combined with the ability of NR to address both structure and the surface excess $\Gamma$ of adsorbed layers. In order to find the ideal photo-switch with not only favorable bulk properties, but also greatly improved light-induced changes in interface chemistry, the interfacial molecular structure, charging state and surface excess need to be resolved and compared. These underlying properties, which are inherited from the photo-switchable surfactant, can be used to tailor colloidal properties of soft hierarchical materials like emulsions and foams and render them active to light stimuli.

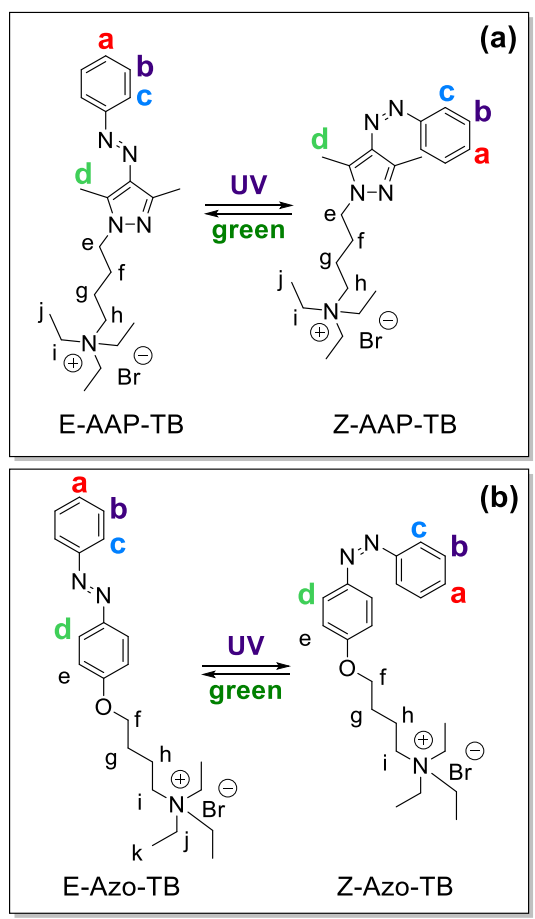

**Figure 1**. *Molecular structures of (a) arylazopyrazole tetraethylammonium bromide (AAP-TB) and (b) azobenzene tetraethylammonium bromide (Azo-TB) surfactants in E and Z conformations. Assignments of the $^1$H NMR signals in Figure 3 employ the nomenclature given by the letters **a** to **j**.*

# EXPERIMENTAL DETAILS

## Sample Preparation

The synthesis protocol for Azo-TB and AAP-TB is described in the Supporting Information (SI). After synthesis, Azo-TB was purified with column chromatography, while the AAP-TB was recrystallized in ethanol several times as described in the SI. Purity was additionally checked using NMR and mass spectroscopy (SI). Solutions of Azo-TB and AAP-TB were prepared by dissolving the surfactants in ultrapure water (18.2 MΩcm; total oxidizable carbon ≤ 3 ppb) which was obtained from a *Merck* Milli-Q Reference A+ purification system. All necessary glassware was pre-cleaned in Alconox solutions (*Sigma Aldrich*), dried and subsequently soaked in concentrated sulfuric acid (98 %, Carl Roth) with Nochromix (Godax Laboratories) for at least 12 h. Subsequently, the glassware was rinsed with copious amounts of ultrapure water. $D_2O$ for the NR experiments was supplied by *Sigma Aldrich* (> 99 % D) All samples were used immediately after preparation and all measurements were performed at 294 K room temperature except for NR and NMR where the measurements were done at 299 K. For *in situ* photo-switching, LEDs with emission wavelengths between 370 and 840 nm were used. The intensity at the sample position was adjusted to ~1.0 mW/cm$^2$ for the 520 nm green LED and to ~7.4 mW/cm$^2$ for the 365 nm UV LED.

## UV/Vis Spectroscopy

Ultraviolet/visible (UV/Vis) absorption spectra were recorded with a *Perkin Elmer* Lambda 650 spectrometer using 10 mm cuvettes. All spectra were recorded directly after the samples were irradiated with light for at least 20 min. To determine the influence of different light irradiation on the absorption spectra, the samples were irradiated first with 365 nm UV light and afterwards with different excitation wavelengths from 370 nm to 840 nm (see Figure S1 for extinction coefficients as a function of wavelengths and Figure S2 for the LEDs normalized emission spectra).

**NMR Spectroscopy**

The $^1$H high-resolution magic-angle spinning (HR-MAS) NMR experiments were conducted on a *Bruker* Avance NEO (11.74 T, $v_L(^1H)$ = 500 MHz) using a 4 mm H/F/X MAS probe equipped with magic-angle gradient coils. All samples were placed in 4.0 mm ZrO$_2$ HR-MAS rotors with an upper spacer made of Teflon and sealed with a Kel-F screw and cap. The active volume inside the rotor was 50 µL. Adamantane was used as an external reference for referencing the chemical shift scale ($\delta(^1H)$ = 1.85 ppm)[47] and optimizing the rf-field strength ($v_{RF}$ = 50.0 kHz and $\tau_{\pi/2}$ = 5.0 µs). All $^1$H HR-MAS NMR spectra were recorded employing 5.0 kHz with a relaxation delay of 30 s. After reaching the spinning rate of 5.0 kHz each sample was left to equilibrate for 5 min. Temperature control and calibration was performed. Two NMR spectra were recorded for all irradiation steps. First, a single pulse $^1$H HR-MAS NMR spectrum, and second a $^1$H HR-MAS NMR spectrum with a low spectral width (3.6 ppm) of the aromatic chemical shift region, were recorded. Thus, the receiver gain could be optimized while suppressing the intense $^1$HOD solvent peak. The recorded $^1$H HR-MAS NMR spectra of the aromatic region are shown in Figures S5 and S6 for AAP-TB and Azo-TB, respectively.

To determine the intensity ratio of both *Z* and *E* isomers, the baseline was corrected with the automatic baseline correction implemented in Topspin 4.0.7. For AAP-TB, the $^1$H signals of both isomers in the aromatic region do not overlap. The aromatic $^1$H signals for the *E* isomer are at higher ppm values (integration region: 7.88 – 7.53 ppm) than for the *Z* isomer (integration region: 7.53 – 7.16 ppm). For the *E* isomer of Azo-TB the aromatic signals are between 8.02 – 7.60 ppm and 7.31 – 7.16 ppm. The signals of the Z isomer are in the range of 7.53 – 7.31 ppm and 7.13 – 6.92 ppm.

The light irradiation of the samples was done *ex situ* with 520 nm green, 365 nm UV or 465 nm blue light for at least 15 min. The samples were quickly filled into the MAS rotor, and after reaching 5 kHz MAS frequency they equilibrated for 5 min to reach a constant temperature. After UV irradiation the sample was split into two fractions. One fraction was immediately irradiated with green light and the other fraction was stored in complete darkness and later irradiated with blue light. This procedure ensures identical starting points for the green and blue light irradiated samples. The Z isomers remained stable while the sample were stored in darkness, as demonstrated by an overnight experiment for AAP-TB (Figure S7).

**Surface Tensiometry**

For dynamic surface tension measurements, a pendant drop tensiometer (DSA100, *Krüss*, Germany) was used where the drop was formed at the end of a syringe cannula and a CCD camera monitored its shape. Through the use of image analysis and the Young-Laplace equation, the surface tension was determined from the drop shape as a function of time until equilibrium was reached. This process was done during continuous irradiation with either 520 or 365 nm light. In order to achieve these irradiation conditions, a modified cuvette was used, which was equipped with two long pass filters (Schott OG590) that had a cut-off wavelength at 590 nm. The application of this filter avoided any influence on the surfactant conformation from the ambient light in the room and the light source from the tensiometer itself. In addition, the cuvette was equipped with two LEDs perpendicular to the optical axis of the tensiometer. The emissions of the LEDs (Roschwege) were at wavelengths of 365 nm and 520 nm (emission spectra in SI), which allowed for continuous irradiation of the pendant drop. The cuvette was filled with a small portion of water in order to saturate the gas phase around the pendant drop and to avoid strong evaporation.

**Vibrational Sum-Frequency Generation Spectroscopy**

SFG spectroscopy is an inherently interface-sensitive method that is based on a second-order non-linear optical effect.[48,49] At the interface of interest, two laser beams, a narrowband picosecond visible (vis) beam with frequency $\omega_{vis}$ and a tunable femtosecond infrared (IR) beam with frequency $\omega_{IR}$, are overlapped in time and in space, generating a third beam with the sum-frequency (SF) $\omega_{SF} = \omega_{vis} + \omega_{IR}$ of the two impinging beams. The intensity of the SF for this second-order process is dependent on the intensities of the two fundamental beams and on the nonresonant $\chi^{(2)}_{Nres}$ and resonant $\chi^{(2)}_{R}$ parts of the second-order electric susceptibility $\chi^{(2)}$.[48,49] At charged interfaces with non-zero double-layer potentials $\phi_0$, a third-order contribution $\chi^{(3)}$ needs to be considered. This contribution is often used to gain qualitative[48,50,51] and in some cases quantitative information[52–54] on the interfacial charging state:

$$I(\omega_{SF}) \propto \left| \chi^{(2)}_{Nres} + \chi^{(2)}_{R} + \frac{\kappa}{\kappa + i\Delta k_z} \chi^{(3)} \phi_0 \right|^2 I_{IR} I_{vis} \qquad (1)$$

In this equation, $\kappa$ is the inverse Debye length and $\Delta k_z$ the wave vector mismatch. The resonant part

$$\chi^{(2)}_{R} = \sum_q \frac{A_q}{\omega_q - \omega_{IR} + i\gamma_q} \qquad (2)$$

is a function of the resonance frequency $\omega_q$ of the q[th] vibrational mode, its bandwidth $\gamma_q$ and of the modes amplitude $A_q$. The amplitude $A_q = \Gamma \langle \beta_q \rangle$ is functions of both the surface excess $\Gamma$ and the orientational average $\langle \cdots \rangle$ of the hyperpolarizability $\beta_q$ of the interfacial molecules. In the isotropic bulk phases, all molecules are in a centrosymmetric environment, and consequently their orientational average is zero due to symmetry reasons. However, interfaces like the air-water interface necessarily break the prevailing bulk symmetry, which leads to nonzero amplitudes and thus, finite SFG intensities. For that reason, SFG spectroscopy is inherently interface selective for materials with inversion symmetry. In this work, SFG spectra were recorded with a homebuilt spectrometer that is described elsewhere.[52] In order to study the broad frequency range of O-H and C-H stretching vibrations (2800 – 3800 cm$^{-1}$), the center frequency of the >300 cm$^{-1}$ femtosecond IR pulse was tuned in 4 steps, while the bandwidth of the narrowband visible pulse at 804.1 nm was <5 cm$^{-1}$. The acquisition time for each IR frequency was between 30 and 60 s and the SFG spectra of the samples were referenced to that of an air plasma-cleaned thin Au film on top of a silicon wafer. Normalization was done in order to remove the frequency-dependent changes in the IR pulse energies. All samples were prepared in a quartz glass petri dish and were irradiated with 520 nm green or 365 nm UV light until equilibrium was reached. The irradiation states were produced by LEDs that were positioned 20 mm above the samples and were equipped with clean-up filters to remove unwanted wavelengths that could possibly interfere with the SFG spectroscopy. The spectra of the applied LEDs are reported in the SI. The times needed to reach equilibrium for the interfaces under investigation were estimated from dynamic surface tension measurements and by recording time-resolved SFG spectra. During both the equilibration procedure and acquisition of SFG spectra, the samples were continuously irradiated with UV/green light. For investigations of the air-water interface, the SF and visible pulses were s-polarized while the IR beam was p-polarized (ssp polarization), and the reference spectrum from the Au film was recorded in ppp polarization.

**Neutron Reflectometry**

The FIGARO instrument[55] at the Institut Laue-Langevin (Grenoble, France) was used to determine $\Gamma$ for both Azo-TB and AAP-TB under both green and UV illumination states with respect to the bulk surfactant concentration using NR. Data of 6 samples in parallel were recorded on a sample changer in kinetic mode,[56] and at least 7 data points were recorded over >100 min for each sample to ensure full equilibration. In each case, the specular neutron reflectivity, R, was recorded with respect to the momentum transfer normal to the interface, Q, defined as

$$Q = \frac{4\pi \sin\theta}{\lambda} \qquad (3)$$

where $\theta$ is the incident angle and $\lambda$ the wavelength. The recently established low-Q analysis method was used,[57] and as such data were recorded only at $\theta = 0.62°$ and in a range of $\lambda = 2 - 16$ Å, and they were reduced in a range of $\lambda = 4.5 - 12$ Å to give the Q-range $0.01 - 0.03$ Å$^{-1}$. All samples were prepared by dissolving the surfactant in air contrast matched water (ACMW), which is 8.1 % by volume D$_2$O in H$_2$O, having a zero scattering length density. To calculate $\Gamma$, the low-Q analysis of a single mixed layer of surfactant tails and headgroups, was used for fitting the data. For thin films at the air-water interface, this low-Q approach has been demonstrated to be virtually independent of the interfacial structure,[58] and hence is effectively model-free. The surface excess $\Gamma$ may thus be calculated as

$$\Gamma = \frac{\rho\tau}{N_A \sum b_i} \qquad (4)$$

where $\rho$ is the scattering length density of the surfactant, $\tau$ is the fitted layer thickness, $N_A$ is the Avogadro number and $b_i$ is the scattering length of each atom $i$. The data were fitted using the Motofit software.[59] The roughness values used in the model were 3 Å. The background was not subtracted from the data, and the value of the background to use in the model was determined from the average of several measurements of neat ACMW recorded for the same acquisition time of 2 min.

For structure information of the surfactants the NR data were recorded at $\theta = 0.62°$ and $3.8°$ in a range of $\lambda = 2 - 30$ Å. The samples where measured in ACMW as well in D$_2$O. For modeling the layer thickness the surfactant was divided into two parts (chain and headgroup) and constraint was applied so that the number of chains and headgroup are equal. From the surface tension the capillary wave roughness was determined. For fitting of the NR result with the surfactants in the Z state (UV light irradiation), we calculated the air mixing in the adsorbate layer instead of determining a fixed volume fraction. For the structure data the background was subtracted from the data. For more details about the fitting procedure we refer the reader to a recent paper on modeling neutron reflectivity data of surfactant monolayers,[60] and full details can be found in the SI. The samples were measured in a 6-position sealed adsorption trough, with modified lids which have allowed us to irradiate the surfactants solutions in the adsorption troughs continuously with green or UV light.

**RESULTS**

**Photo-response and properties in the bulk solution**

In Figure 2a and 2b, UV/Vis absorbance spectra of 0.1 mM AAP-TB and Azo-TB aqueous solutions that were irradiated with 520 nm green and 365 nm UV light are presented. In order to determine the position of the absorbance bands more accurately we have fitted the UV-Vis spectra using Gaussian functions and report the band positions in Table 1.

**Table 1.** Band positions and their assignments [a)] in UV/Vis absorbance spectra of AAP-TB and Azo-TB in *E* and *Z* conformations due to irradiation with 520 nm green and 365 nm UV light, respectively. [33,34,61,62]

|  | *E* | | *Z* | |
| --- | --- | --- | --- | --- |
|  | π → π* | n → π* | π → π* | n → π* |
| **AAP-TB** | 330 nm | 423 nm | 300 nm | 432 nm |
| **Azo-TB** | 342 nm | 433 nm | 308 nm | 433 nm |

[a)] For that spectra in Figures 1a-b have analyzed using Gaussian fits to the spectra.

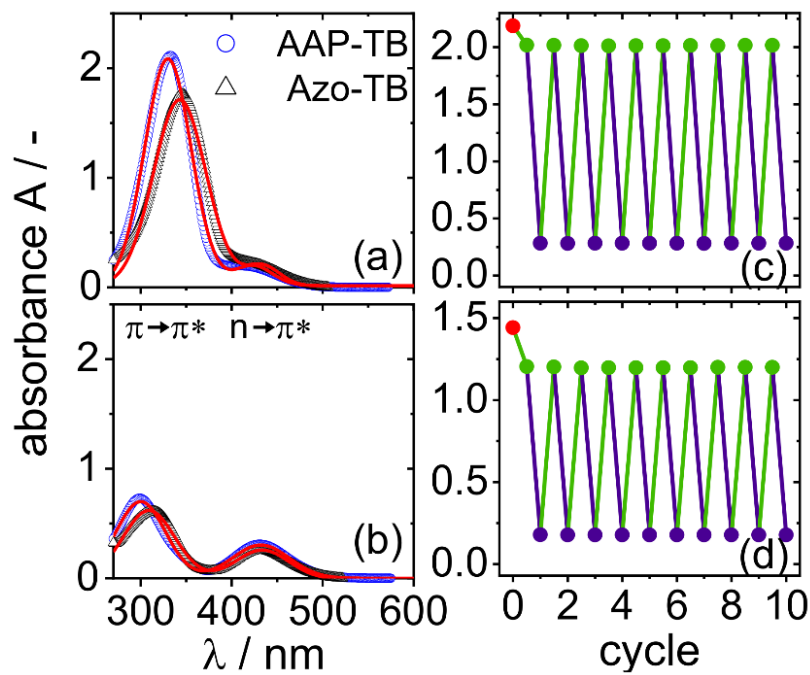

**Figure 2.** *UV/Vis absorbance spectra of 0.1 mM AAP-TB and 0.1 mM Azo-TB in aqueous solution after irradiation with (a) 520 nm green and (b) 365 nm UV light. The spectra were fitted with Gaussian functions (red lines) to determine the position of the n $\rightarrow \pi^*$ and $\pi \rightarrow \pi^*$ transitions. Note that the scales of the y-axis for (a) and (b) are equal. Equilibrium absorbance of the band at 333 nm for (c) AAP-TB and at 346 nm for (d) Azo-TB for 10 switching cycles between UV (blue circles) and green light (green circles), respectively. The red circles in (c) and (d) represent the absorbance at 333 nm for samples in the overall dark state, which was established by storing the samples for 3 weeks in the dark. Full spectra are presented in the SI (Figure S3). Lines in (c) and (d) guide the eye and are simply connecting the data points between green and UV light irradiation for each irradiation cycle.*

When the samples were irradiated with green light, a strong π → π* band that was centered at a wavelength of 330 nm for AAP-TB and at 342 nm for Azo-TB was observed.[61–63] In addition to this band, a much weaker n → π* band centered at ~423 nm for AAP-TB and 433 nm for Azo-TB was noticeable in the absorbance spectra. For samples that were subjected to 365 nm UV light, the absorbance of the π → π* band decreased and was blue-shifted to shorter wavelengths, whereas the n → π* band increased in absorbance and was red-shifted to longer wavelengths for both surfactants (Table 1). Clearly, the above described differences in the absorbance spectra are indicative for photo-isomerization reactions from the *E* (green light) to *Z* (UV light) conformations of each surfactant. The results in Figure 2 and Table 1 also show that the π → π* band of AAP-TB is blue shifted compared to Azo-TB. This is different for the n → π* band, which is at similar wavelengths when the UV/Vis spectra of the two surfactants are compared. This observation clearly indicates a larger splitting of π → π* and n → π* bands for AAP-TB compared to Azo-TB, a fact that is corroborated by previous reports on other AAP derivatives.[33,34] The larger splitting of the of π → π* and n → π* bands can also result to more favorable photo-switching properties as we will discuss in detail below.

In order to showcase, the photo-stability of AAP-TB and Azo-TB we present in Figure 2c and 2d the absorbance of the π → π* bands for AAP-TB and for Azo-TB after the samples were in equilibrium but for several E/Z switching cycles where the samples were first irradiated with green light, equilibrated and subsequently irradiated with UV light and equilibrated again. Clearly, the absorbance of the π → π* band is not changing between cycles which is indicative for reversible switching and a high photo stability of both surfactants. In addition, experiments where the excitation wavelength was tuned between 370 and 840 nm (see SI) were performed, where samples were first irradiated with 365 nm UV light and after reaching equilibrium a back-switch experiment with different excitation wavelengths was performed. The results of these experiments are presented in Figure S1, where the ratio of the extinction coefficient between the *E* isomer and the *Z* isomer (365 nm) are shown. From a close analysis of these data, we conclude that both AAP-TB and Azo-TB show the highest *Z*→*E* switching ability for an excitation wavelength of 520 nm.

In Figure 3, we present $^1$H NMR spectra of both surfactants and report on the changes in $^1$H chemical shift and their intensities for the *E* and *Z* isomers of AAP-TB and Azo-TB resulting from *ex situ* irradiation experiments in Table 2 and Figure S5 and S6. At thermal equilibrium, AAP-TB includes both isomers with distinct and well separated $^1$H chemical shifts with prevalence of 87 % *E* and 13 % *Z*; Figure 3a. After irradiation with green light (520 nm), the prevalence of the *E* isomer (93 %) clearly increases; Figure 3b. Subsequent UV irradiation (365 nm) leads to a conformational change and the *Z* isomer is more prevalent (89 %). Switching back with green light restores the initial state of 90 % *E*. Note that blue light irradiation (465 nm) is not as effective as UV light and only yields 78 % *E* isomer (Figure S5e). For Azo-TB, after the first irradiation 84 % *E* and 16 % *Z* isomers are present in the sample. The subsequent UV irradiation is very efficient and results in 3 % *E* and 97 % *Z* isomer. Again, the switching with green light (82 % *E*) is more efficient than switching with blue light (67 % *E*; Figure S6d). The amounts of *E* isomers for AAP-TB and Azo-TB after the different irradiation procedures is summarized in Table 2. To test the thermal stability of the *Z* isomer of AAP-TB, a series of $^1$H HR-MAS experiments were recorded over 16 h after an irradiation with UV light, where within the 16 h under dark conditions no significant intensity changes for the *E* and *Z* isomers (90 % *Z* vs. 87 % *Z*) were found (Figure S7).

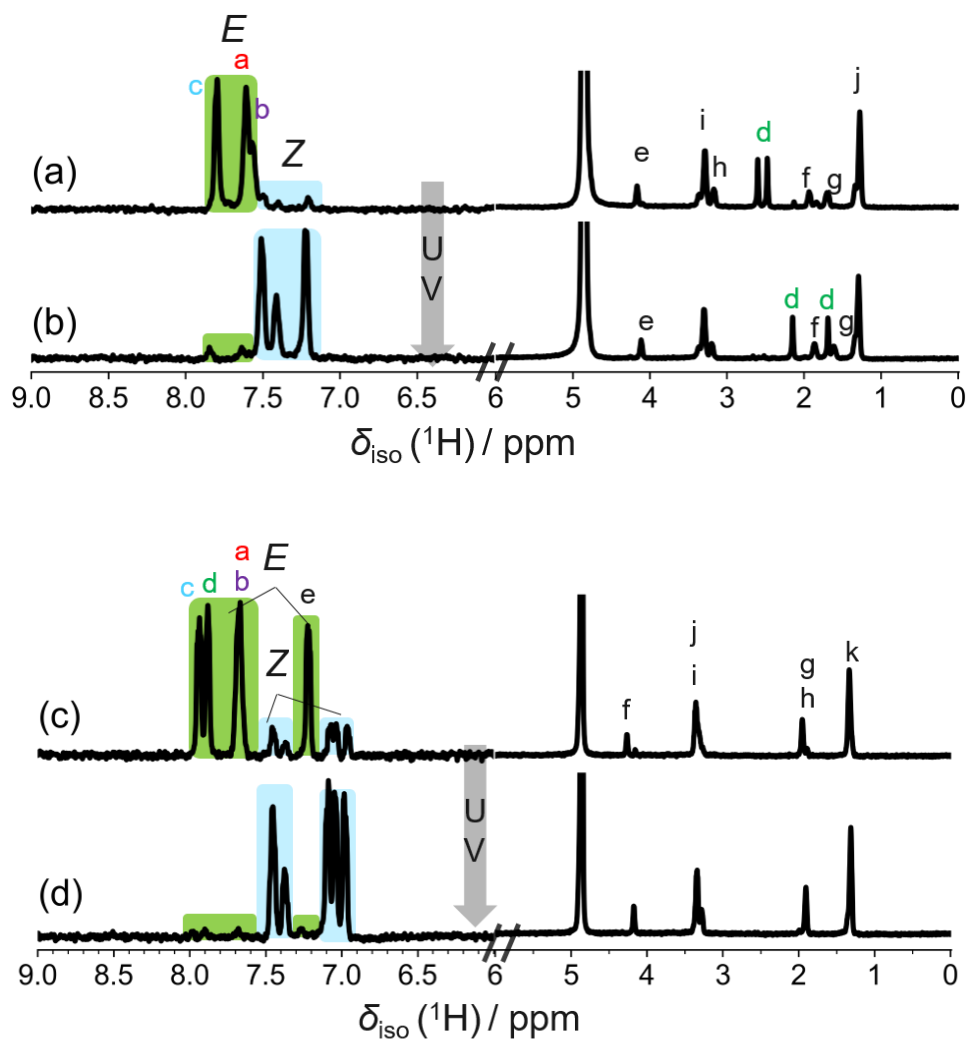

**Figure 3** $^1$H HR-MAS NMR spectra of 7 mM (a) E-AAP-TB, (b) Z-AAP-TB, (c) E-Azo-TB and (d) Z-Azo-TB recorded immediately after (a/c) 520 nm green irradiation showing predominantly the E isomer and (b/d) 365 nm UV illumination showing predominantly the Z isomer. All experiments were recorded at 11.74 T and T = 299 K employing 5.0 kHz MAS. The assignment of $^1$H signals follows that of Figure 1.

In addition to NMR spectroscopy, we determined the PSS from UV/Vis spectroscopy and present the details of our analysis in the SI. The results from UV/Vis are fully consistent with the NMR data as we find for AAP-TB 92 % of the *E* isomer when irradiated by green light and 5% of the *E* isomer under UV light, while we observed 82 % and 5 % for the *E* isomer of Azo-TB when irradiated with green and UV light, respectively. We want to point out that the PSS from UV/Vis spectroscopy correspond to a reduced concentration of 0.1 mM to keep the samples maximum absorbance within the detection limit of our spectrometer. Our analysis of the maximum Z→E switchability in the bulk using UV\Vis spectroscopy is additionally corroborated by *ex situ* irradiation experiments of AAP-TB and Azo-TB followed by $^1$H HR-MAS NMR experiments using different wavelengths and have demonstrated that 520 nm green light is more efficient for Z→E switching for both surfactants compared to 465 nm blue light (see Table 2). Moreover, AAP-TB shows a higher photostationary state (PSS) with respect to Z→E switching under green light irradiation compared to Azo-TB, whereas the PSS for E→Z switching under UV light is higher for Azo-TB.

**Table 2.** *Photostationary states (PSS) from NMR intensities and UV/Vis for the E isomer of AAP-TB and Azo-TB surfactants under different light irradiation conditions (green = 520 nm; blue = 465 nm, UV=365 nm), and the maximum changes in equilibrium surface tension $\Delta\gamma_{max}$, the maximum change equilibrium in surface excess $\Delta\Gamma_{max}$ from NR, and the maximum change in SFG amplitude of the aromatic C-H band $\Delta A_{q\ max}$ for AAP-TB and Azo-TB below their CMCs.*

| Surfactant (irradiation)[a] | E / % (NMR) | E / % (UV\Vis) | $\Delta\gamma_{max}$ / mN/m | $\Delta\Gamma_{max}$ / μmol/m$^2$ | $\Delta A_{q\ max}$ (arom. C-H) |
|---|---|---|---|---|---|
| **AAP-TB** (green) | 93 ± 3 | 92 ± 2 | | | |
| **AAP-TB** (blue) | 78 ± 3 | 75 ± 2 | 12 ± 0.9 | 2.15 ± 0.07 | 0.55 ± 0.06 |
| **AAP-TB** (UV) | 11 ± 3 | 5 ± 2 | | | |
| **Azo-TB** (green) | 84 ± 2 | 82 ± 2 | | | |
| **Azo-TB** (blue) | 67 ± 2 | 65 ± 2 | 8 ± 0.5 | 1.02 ± 0.03 | 0.21 ± 0.02 |
| **Azo-TB** (UV) | 3 ± 2 | 5 ± 2 | | | |

**Light-induced interfacial changes**
**Surface tension**

In Figure 4, the equilibrium surface tension isotherms of AAP-TB and Azo-TB are presented with respect to the light irradiation (green vs. UV). There are consistently higher surface tension values when the samples were irradiated with UV light than green light. The surfactant critical micelle concentration (CMC), as indicated by the apparent kinks in the surface tension isotherms, is different for the two light irradiations for AAP-TB and is shifted from 15 to 30 mM for green and UV light, respectively. This is different for Azo-TB, where the CMC of 20 mM was independent of the light irradiation. We will now discuss the maximum changes in equilibrium surface tension $\Delta\gamma_{max}$, which we define at a given concentration as the difference $\Delta\gamma$ between the equilibrium surface tensions under UV and green light irradiation ($\Delta\gamma = \gamma_{UV} - \gamma_{gr}$). $\Delta\gamma$ is nonzero for both surfactants and positive when the concentration was between 1 and 10 mM, where a maximum difference $\Delta\gamma_{max}$ of ~12 mN/m and ~8 mN/m was observed for AAP-TB and Azo-TB, respectively. This result indicates a slightly higher switching ability below the CMC in terms of surface tension changes at the air-water interface for AAP-TB compared to Azo-TB. The solid lines in Figures 4a and 4b show fits to the surface tension isotherms using a Frumkin adsorption isotherm for a surfactant monolayer adsorbed at the air-water interface:

$$bc = \frac{\theta}{1-\theta} e^{-a\theta} \qquad (5)$$

and the equation of state

$$\gamma = \gamma_0 + \Gamma_{max} RT \left[\ln(1-\theta) + a\theta^2\right] \qquad (6)$$

where $b$ is the adsorption equilibrium constant for surfactants at the air-water interface, $c$ the surfactant concentration in the bulk solution, $\theta = \Gamma/\Gamma_{max}$ is the surface coverage and *a is* an interaction parameter. In addition, $\Gamma$ and $\Gamma_{max}$ are the surface excess and the limiting surface excess at monolayer coverage at the air-water interface, respectively. For more information about this model, we refer the reader to the works of Miller and co-workers[64,65] and to the supporting information for the free parameters in our fitting procedures. For fitting the isotherms, we used as fixed parameters the values of $\Gamma_{max}$ resolved using the model-free approach of NR, the results of which are discussed in more detail below. The values of $\Gamma_{max}$ were resolved as 4.4 and 3.5 µmol/m² for AAP-TB for predominant *Z* and *E* conformations, respectively, and 3.3 µmol/m² for Azo-TB under green light (*Z* state). As a result of fixing $\Gamma_{max}$ in the Frumkin model from the NR data, we were able to achieve excellent agreement with the surface tension data for the surfactants being in the *E* conformations and for the *Z* conformation of AAP-TB. However, the Frumkin model for surfactants in the *Z* conformation for Azo-TB was inadequate to reproduce the experimental data, when the $\Gamma_{max}$ was fixed to the value of 3.3 µmol/m² which is the model-free result from NR. As a consequence, it was necessary to assume $\Gamma_{max}$ as an additional free fitting parameter in order to achieve a reasonably good fitting result to the surface tension isotherm in Figure 4b. This has resulted in better fitting, but also in a dramatically higher $\Gamma_{max}$ of 11.5 µmol/m². In fact, this value is more than three time higher as the value (3.3 µmol/m²) from NR.

Furthermore, we did a structure analysis with NR for Azo-TB at a concentration of 10 mM. The data are presented in the SI. For an analysis of the reflectivity curves, Azo-TB was divided into two parts, a hydrophobic chain and a hydrated hydrophilic head group (Figure S10). For the surfactants in *E* conformations (green irradiation) the fitted thickness of the adsorbed surfactant layer was 5.8 Å, whereas under UV light the thickness was 5.1 Å.

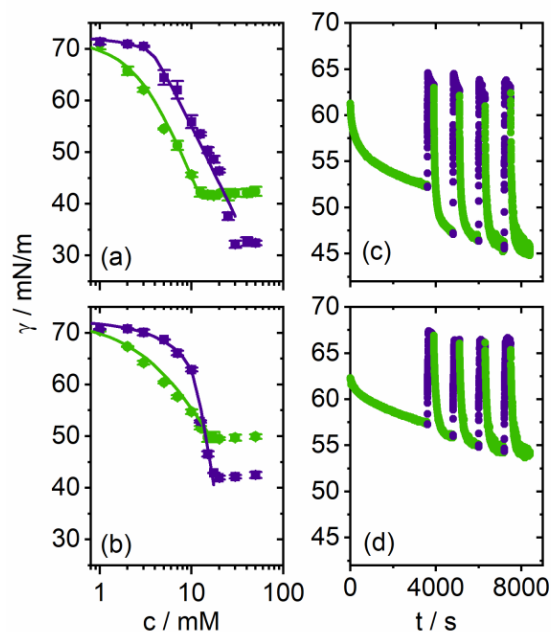

**Figure 4.** *Equilibrium surface tension values of (a) AAP-TB and (b) Azo-TB solutions as a function of bulk concentration and with respect to the light illumination for 520 nm green (green circles) and 365 nm UV (dark blue squares), which are dominated by the E and Z isomers, respectively; see Figure 1. Solid lines represent fits to the surface tension data using a Frumkin adsorption model, which is described in the main text. Dynamic surface tension measurements of E/Z and Z/E photo-switching cycles of 7 mM (c) AAP-TB and (d) Azo-TB solutions; these concentrations correspond to the highest changes in surface tension with light irradiation at concentrations below the cmc.[51]*

In order to further assess the difference between AAP and Azo surfactants we also performed kinetic measurements, where the time-dependent changes upon $E{\rightarrow}Z$ and $Z{\rightarrow}E$ photo-isomerization were investigated with surface tensiometry. The results of these experiments are shown in Figure 4c and 4d, where dynamic surface tension measurements of 7 mM AAP-TB and 7 mM Azo-TB are presented for irradiation with green and UV light.

Analysis of the dynamic surface tensions in Figures 4c and 4d using first-order kinetics[51] yields rate constants of 0.07 s$^{-1}$ and 0.02 s$^{-1}$ when switching from the $E$ to the $Z$ state for AAP-TB and Azo-TB, respectively (Figure S4). The faster kinetic for AAP-TB is not observed for the $Z{\rightarrow}E$ back-switch using green light. Here, the switching kinetics for both surfactants are slower and similar rate constants of 0.01 s$^{-1}$ are observed. We would like to point out that this is the first comparison of the switching kinetics and non-equilibrium interfacial properties for AAP vs. Azo amphiphiles at a fluid interface. Clearly the switching dynamics are slightly more favorable for the arylazopyrazole compared to the azo amphiphile and will be discussed below.

**Surface excess and molecular order**

Figure 5 presents SFG spectra of AAP-TB and Azo-TB at the air-water interface for irradiation with green as well as UV light. Before we discuss the major changes in the presented SFG spectra as a function of concentration we will describe the overall shape of the spectra and the main vibrational bands that can be observed. For that we recall that the SFG intensity originates from resonant contributions to the second-order electric susceptibility due to vibrational resonances of interfacial molecules and their non-resonant $\chi^{(2)}_{Nres}$ contribution; for a discussion of SFG principles, the reader is referred to the experimental details section.

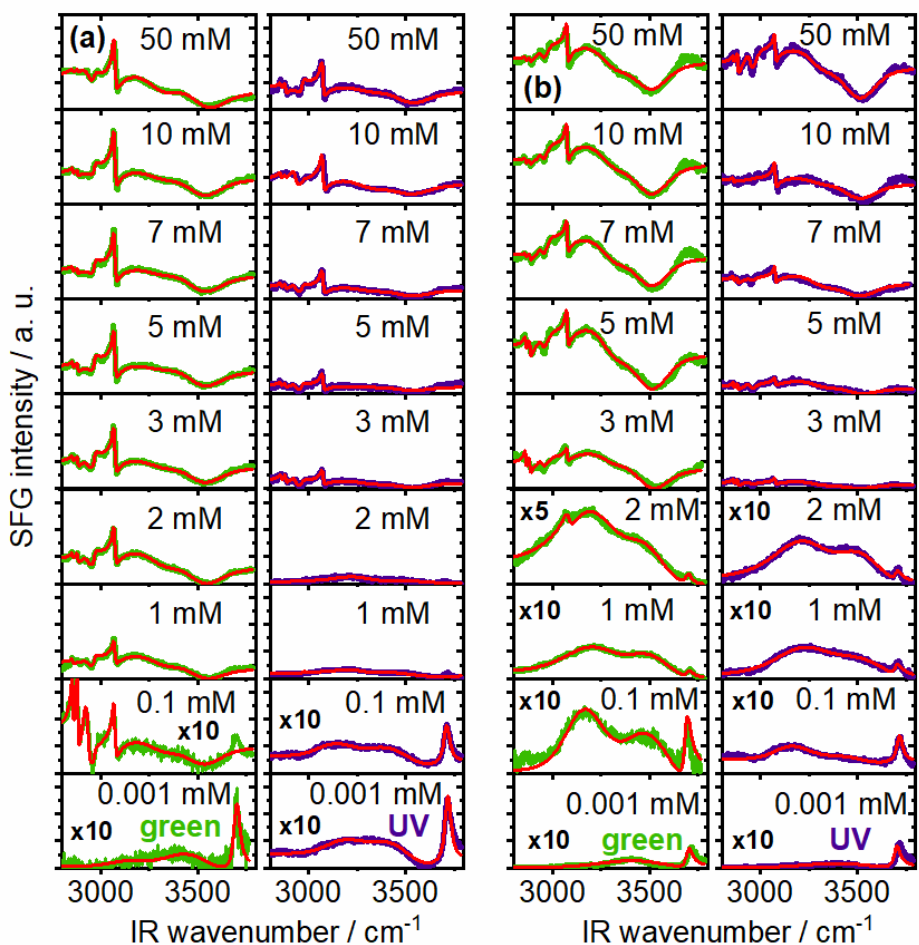

**Figure 5.** *Vibrational SFG spectra at the air-water interface of (a) AAP-TB and (b) Azo-TB solutions at different concentrations and under different light irradiation conditions for 520 nm green and 365 nm UV, which are dominated by the E and Z isomers, respectively; see Figure 1. The solid red lines represent nonlinear least square fits to the experimental data, as explained in the text. Spectra for the low concentration for AAP-TB and Azo-TB were multiplied by a factor of 10.*

While $\chi^{(2)}_{Nres}$ is often very small, there can be substantial $\chi^{(2)}_{Nres}$ contributions to the SFG intensity in cases where either of the visible or SF frequencies is close to optical transitions of the molecules under investigation. For instance, metal surfaces like Au[66,67] or other interfaces that have been modified with dye molecules[68,69] or photoswitches[51,70] show strong $\chi^{(2)}_{Nres}$ contributions that can heterodyne the resonant $\chi^{(2)}_R$ signals due to the coherent overlap of $\chi^{(2)}_R$ and $\chi^{(2)}_{Nres}$ contributions (see equation (1)), as has been discussed in detail by Yang and Hore.[71] For example, optical transitions of the dye Rhodamine 6G are around 532 nm, and as was previously shown by Wang and co-workers[69] can lead to substance double resonance effects if similar wavelengths for the SFG or the visible pulses are chosen. In the present case of AAP-TB and Azo-TB at the air-water interface we have no direct optical resonances in our range of wavelength but with the SFG wavelengths in our experiments between 650 to 620 nm we are close to the optical transition of the photo-switches. For that, we recall that the best optical excitation to trigger E/Z conformational changes is achieved by excitation with 520 nm (Table 2, Figure S1). Although, we do not expect a double resonance for AAP-TB and Azo-TB, clearly, their $\chi^{(2)}_{Nres}$ contributions are rather strong in our experiments but show negligible frequency dependences within our experimental range of wavelengths. In fact, close analysis of the SFG spectra in Figure 5 shows that at high bulk surfactant concentrations the substantial $\chi^{(2)}_{Nres}$ contribution causes highly dispersive line shapes of the resonant vibrational bands and leads to an apparent frequency-independent offset in the spectra. The dispersive line shape of the vibrational bands is most clearly seen by analyzing the sharp vibrational band at 3070 cm$^{-1}$ due to aromatic C-H stretching[50,72] vibrations and the broad and overlapping bands centered at 3200 and 3450 cm$^{-1}$, which are attributed to O-H stretching vibrations[73,74] of interfacial hydrogen-bonded water molecules. Further resonances[70,75] of the CH$_2$ (~2855 cm$^{-1}$), the CH$_3$ (~2880 cm$^{-1}$) symmetric stretching as well as the CH$_3$ antisymmetric stretching (2965 cm$^{-1}$) modes and the CH$_3$ Fermi resonance (2930 cm$^{-1}$) can be also inferred from the SFG spectra in Figure 5. However, in the following we concentrate on the aromatic C-H as well as on the O-H stretching bands and $\chi^{(2)}_{Nres}$ contributions.

In addition to the SFG amplitudes of the specific contributions mentioned above, the overall shape of the SFG spectra presented in Figure 5 is also of great interest. The shape can most clearly be elaborated by addressing the SFG spectra from 50 mM AAP-TB and 50 mM Azo-TB solutions. The presence of these cationic surfactants causes an excess of positive charges at the air-water interface and leads to polarization and ordering of interfacial water molecules within the interfacial electric field and strong O-H bands can be observed.[73,76] As already noted above the SFG spectra show a strong nonresonant contribution $\chi^{(2)}_{Nres}$ which can heterodyne both O-H and C-H bands and can lead to a reduction of the O-H intensity in the SFG spectra depending on the interference conditions (constructive vs. destructive). Nonlinear least-square fitting of the SFG spectra with model function according to equation (1) shows that excellent fitting results could be achieved (see Figure 5), but only when a relative phase of π between the $\chi^{(2)}_{Nres}$ and the resonant contribution $\chi^{(2)}_{R,OH}$ from the O-H stretching mode of interfacial water molecules was assumed. More details about the fitting results and the parameters used in our fit procedures can be found in the Supporting Information. Close inspection of the changes in $\chi^{(2)}_{Nres}$, the aromatic C-H band at 3070 cm$^{-1}$ as well as of the O-H bands clearly shows that these contributions increase with the surfactant concentration. $E \rightarrow Z$ photo-isomerization also results in a decrease in SFG intensity of all contributions for concentrations <10 mM (Figure 5). In some cases, this change was quite drastic, e.g., for AAP-TB surfactant at a concentration of 1 mM, where a spectrum that resembled fully resolved surfactant with strong aromatic C-H and O-H bands is converted to one that closely resembles a neat air-water interface (in the absence of surfactant). In this case, an additional narrow band at 3700 cm$^{-1}$ can be noticed and is attributable to free so-called dangling O-H groups that have no hydrogen bonds and point into the gas phase.[73,77] Clearly, this transition can be taken as a first evidence that the surface excess is drastically changing with light irradiation, as was already suggested by surface tensiometry (Figure 4). As a consequence of the changes in surface excess, the charging state of the air-water interface decorated with AAP-TB or Azo-TB surfactants is drastically changing, this results in similar drastic change in the O-H contribution because the latter are coupled to the double-layer layer potential $\phi_0$. For that we refer to equation (1) in the SFG details section above. In Figure 6 (left axes), the nonresonant contribution $\chi^{(2)}_{Nres}$ and the amplitude $A_{C-H}$ of the aromatic C-H band at 3070 cm$^{-1}$ are shown for AAP-TB and Azo-TB solutions as a function of concentration with respect to the illumination condition (green vs. UV). Also, in Figure 6 (right axes), values of $\Gamma$ from the low $Q$-analysis of NR are presented. The two sets of data are scaled so that points from the two techniques (SFG and NR) overlap above the CMCs in order for us to identify any deviations in their proportionality below the CMCs. It is clear that trends in the data of $\chi^{(2)}_{Nres}$, which only depends on the surface excess, from SFG spectroscopy in panels (a) and (c) of Figure 6 coincide very well with those of $\Gamma$ from NR for both AAP-TB and Azo-TB, respectively, i.e., the data are roughly proportional. However, through a comparison of the data of $\chi^{(2)}_{Nres}$ (depicted as a red solid line) with the amplitude of the aromatic C-H band $A_{C-H}$ and the surface excess $\Gamma$ from NR in Figures 6b and 6d, we observe in Figure 6b a strong deviation in the trends only in the case of AAP-TB under green irradiation with the surfactants predominantly in the $E$ conformation. Because the SFG amplitude $A_{C-H} = \Gamma \langle \beta_{C-H} \rangle$ is a function of $\Gamma$ and the orientational average $\langle \cdots \rangle$ of the molecular hyperpolarizability $\beta_{C-H}$, we can conclude that there must be a change in the orientational average $\langle \beta_{C-H} \rangle$ of this surfactant with decreasing surface coverage, i.e., a change in interfacial molecular order. Clearly, this is different for Azo-TB, where the relative values of $A_{C-H}$, $\chi^{(2)}_{Nres}$ and $\Gamma$ compared with their maxima are roughly proportional over the full concentration range, which is discussed in more detail below.

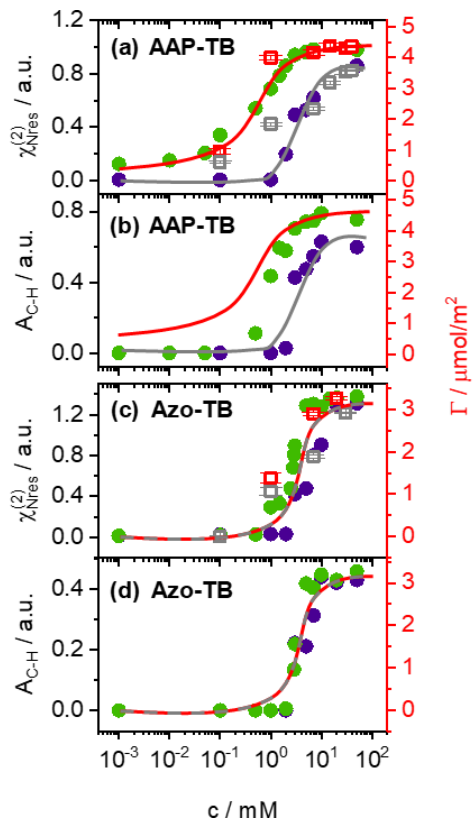

**Figure 6.** *Magnitude of the nonresonant contributions $\chi^{(2)}_{Nres}$ (filled circles; left axes) for (a) AAP-TB and (c) Azo-TB as well as amplitude $A_{C-H}$ of the aromatic C-H stretching band (filled circles; left axes) for (b) AAP-TB and (d) Azo-TB solutions as a function of the bulk concentration and with respect to the illumination state: symbols in green and dark blue color indicate green (predominant E conformation) and UV illuminations (predominant Z conformation), respectively. Solid lines in (a) and (c) are a guide to the eye, which are reproduced exactly in (b) and (d). The equilibrium surface excess from NR as a function of the concentration and with respect to the illumination state is shown in (a) and (c) (right axes), where open red and grey squares indicate green and UV irradiation, respectively.*

**DISCUSSION**

First, we want to discuss the differences in the behavior of AAP-TB and Azo-TB surfactants in the bulk solution. Our results from NMR spectroscopy demonstrated that the PSSs for AAP-TB and Azo-TB are very similar, although not identical: under green irradiation, AAP-TB had a PSS of 93 % for the *E* isomer compared to Azo-TB where the PSS was at 84 %. In contrast, the PSS of Azo-TB was 97 % for the *Z* isomer when irradiated with UV light, while the PSS of AAP-TB was 89 %. In addition, we have shown that the switching to the *E* isomer with blue light leads to a much lower PSS as for green light irradiation. At this point, we would like to note that in previous works,[33,34] where a comparison of arylazopyrazole and azobenzene derivatives the bulk was reported, the authors did not explicitly report the influence of the light wavelength in detail. In fact, in these works the PSS of the AAP derivatives (>90 %) was investigated using green light irradiation, while the PSS of azobenzene derivatives (70-80 %) was determined with blue light irradiation. However, Zhu et al. analyzed the influence of the light wavelength and reported for the azobenzene derivatives in their study the best switching ability with 365 nm UV (*E*→*Z*) and 656 nm red light (*Z*→*E*).[78]

Although the behavior of AAP-TB and Azo-TB surfactants in the bulk solution was very similar, there are substantial differences at the air-water interface as will discuss below. Here our experimental results – resulting from several complementary techniques – show larger changes for *E*→*Z* photo-isomerization for AAP-TB than for Azo-TB. This can be seen through a comparison of the limiting surface excess $\Gamma_{max}$, which we have obtained from a model-free analysis of neutron reflectivity data at the air-water interface. The values for AAP-TB change from ~4.4 to ~3.5 µmol/m$^2$ when going from the *E* to the *Z* state, while the corresponding values for Azo-TB change only from ~3.3 to ~2.9 µmol/m$^2$. The more pronounced changes in the surface excess for AAP-TB is likely due to the AAP center, which has a bulky tail with the two methyl groups at the pyrazole ring and therefore requires more space at the air-water interface when it is predominantly in the *Z* conformation However, $\varGamma_{max}$ is higher for AAP-TB than for Azo-TB even in its *Z* state. This suggests that AAP-TB surfactants are able to form closed-packed monolayers when the molecules are predominantly in the *E* conformation. After switching to UV light (*E*→*Z*) the surfactants desorb partly from the air-water interface because of the steric effects, but are still packed in a monolayer. For Azo-TB, on the other hand, the adsorbed monolayers are somewhat disordered under green light irradiation even at high concentration, when $\varGamma_{max}$ is reached, which we infer of the lower surface excess as for AAP-TB.

The structure data from NR bring some support to the above conclusion: For a fully stretched out hydrophobic chain, we estimate a theoretical maximum length of 9.5 Å, whereas the fitted thickness of the adsorbed surfactant layer in the predominant *E* state was only 5.8 Å with a chains volume fraction of 1. This physical picture is consistent with our earlier observation that Azo-TB forms rather open and disordered structures at the air-water interface. Under UV light irradiation, a value of 5.1 Å for the thickness of the chains layer was fitted with a volume fraction of 1. This is in good agreement with the observed surface excess from NR, which only changes slightly between Azo-TB in their *E* and *Z* states. We can conclude that with switching to UV light $\varGamma$ and the thickness for Azo-TB show a small decrease, which means that the surfactants switch to the *Z* isomer and rearrange slightly at the air water interface.

Next, we want to discuss the concentration dependent changes for AAP-TB and Azo-TB. In comparison with the results from SFG spectroscopy and NR for bulk concentrations below the CMC of each surfactants, we observed that the SFG amplitude of the aromatic C-H band underestimates the trend in $\varGamma$ from NR only in case of AAP-TB, while the data roughly coincide for Azo-TB. We recall, that the SFG amplitude $A_q = \Gamma \langle \beta_q \rangle$ depends on the surface excess $\Gamma$ of the surfactant and its orientational average of $\langle \beta_q \rangle$. The different behavior of the SFG amplitude from aromatic C-H stretching vibrations as compared to the changes in surface excess from NR is, thus, clearly related to changes in molecular order of the interfacial layer of AAP-TB, whereas for the influence of molecular order on the SFG amplitude of the aromatic C-H band from Azo-TB is small and in this case the amplitude is dominated by the surfactants surface excess.

**CONCLUSIONS**

Photo-switchable arylazopyrazole tetraethylammonium bromide (AAP-TB) and azo tetraethylammonium bromide (Azo-TB) amphiphiles were studied on a molecular level. In particular, their use as photo-responsive building blocks was addressed and the performance of the two different photoswitches – Azo vs. AAP – has been compared in both aqueous bulk solution (water) and at the air-water interface. In the bulk water, both surfactants show the highest photostationary state (PSS) for switching from the *E* to the *Z* isomer that is when they were irradiated with 520 nm green light, while the back switch from the *Z* to the *E* isomer was initiated by irradiating the samples with 365 nm UV light. The switchability in the bulk is very similar for both surfactants, and the AAP-TB has a higher PSS for the *E* isomer under green light, whereas the Azo-TB showed the higher PSS for the *Z* isomer under UV light. At the air-water interface, we resolved at equilibrium conditions the surface tension as well as quantitatively the surface excess from neutron reflectometry (NR), while the surface excess and the interfacial molecular order was qualitatively addressed from vibrational sum-frequency generation (SFG) where we have analyzed the surfactants resonant contributions from aromatic C-H stretching vibrations and their non-resonant contributions to the second-order electric susceptibility. We observed pronounced changes for AAP-TB surfactant in terms of surface tension, surface excess and changes in SFG spectra when the molecules undergo $E{\rightarrow}Z$ and $Z{\rightarrow}E$ photo-isomerization reactions, while differences were only modest for the Azo-TB. The interfacial responsiveness is clearly superior of AAP-TB surfactants to the weakly responsive behavior of Azo-TB moieties. We point out that both surfactants were newly designed for this study and had very similar molecular structures except for their photoactive unit (AAP vs. Azo). The better performance at the interface of the AAP-TB is likely related to its steric properties, which makes this surfactant interesting for applications in photo-responsive materials and interfaces.

**ASSOCIATED CONTENT**

The Supporting Information includes additional UV/Vis spectra for different irradiation wavelengths (370 – 840 nm), dynamic surface tension measurements, normalized emission spectra of all LEDs used in this work, supporting NMR spectra as well as a complete description of the synthesis protocol for AAP-TB and Azo-TB surfaces and the corresponding product analysis.

**ACKNOWLEDGMENT**

The authors gratefully acknowledge funding from the European Research Council (ERC) under the European Union's Horizon 2020 research and innovation program (Grant Agreement 638278). The authors thank the ILL for allocations of neutron beam time on FIGARO (DOI: 10.5291/ILL-DATA.9-12-540 and 10.5291/ILL-DATA.9-10-1573) and Simon Wood for technical assistance during the experiment. We also thank the Deutsche Forschungsgemeinschaft (SFB 858) for financial support.

**TOC GRAPHIC**

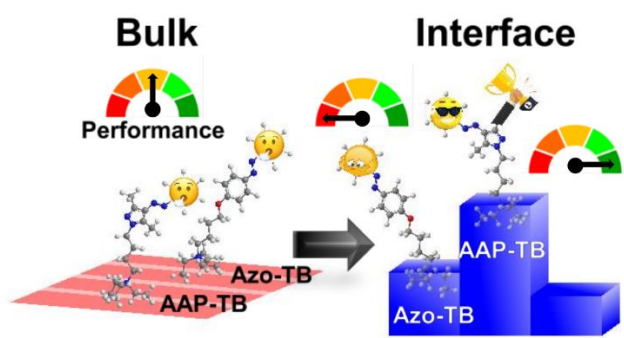